\documentclass[a4paper,amsmath,amssymb,10pt,aps,pre,twocolumn,floats,floatfix,superscriptaddress,showpacs]{revtex4}

\usepackage[T1]{fontenc}
\usepackage{inputenc}
\usepackage{graphicx}
\usepackage{hyperref}

\usepackage{dcolumn}
\usepackage[english]{babel}
\usepackage{times}
\usepackage{color}
\usepackage{psfrag}

\newcommand{\eq}{\! = \!}
\newcommand{\cv}{CV}

\begin{document}

\title{Epidemic variability in complex networks}

\author{Pascal Cr{\'e}pey}
\affiliation{INSERM, Unit{\'e} de Recherche ``{\'E}pid{\'e}miologie, Syst{\`e}mes d'Information 
et Mod{\'e}lisation'' (U707), Paris, F-75012, France}
\affiliation{Universit{\'e} Pierre et Marie Curie, Facult{\'e} de M{\'e}decine Pierre et 
Marie Curie, Unit{\'e} de Recherche ``{\'E}pid{\'e}miologie, Syst{\`e}mes d'Information et 
Mod{\'e}lisation'', Paris, F-75012, France}
\author{Fabi\'an P. Alvarez}
\affiliation{INSERM, Unit{\'e} de Recherche ``{\'E}pid{\'e}miologie, Syst{\`e}mes d'Information 
et Mod{\'e}lisation'' (U707), Paris, F-75012, France}
\affiliation{Universit{\'e} Pierre et Marie Curie, Facult{\'e} de M{\'e}decine Pierre et 
Marie Curie, Unit{\'e} de Recherche ``{\'E}pid{\'e}miologie, Syst{\`e}mes d'Information et 
Mod{\'e}lisation'', Paris, F-75012, France}
\author{Marc Barth\'elemy}
\affiliation{CEA-Centre d'Etudes de
  Bruy{\`e}res-le-Ch{\^a}tel, D\'epartement de Physique Th\'eorique et
    Appliqu\'ee\\
BP12, 91680 Bruy\`eres-Le-Ch\^atel, France}
\affiliation{School of Informatics, Indiana University, Eigenmann Hall,\\ 
Bloomington, IN 47408, USA}

\date{\today}
\widetext

\begin{abstract}
\vspace{0.2cm}
We study numerically the variability of the outbreak of diseases on
complex networks. We use a SI model to simulate the disease spreading
at short times in homogeneous and in scale-free networks. In both
cases, we study the effect of initial conditions on the epidemic
dynamics and its variability.  The results display a time regime
during which the prevalence exhibits a large sensitivity to noise. We
also investigate the dependence of the infection time of a node on its
degree and its distance to the seed. In particular, we show that the
infection time of hubs have non-negligible fluctuations which limit
their reliability as early-detection stations. Finally, we discuss the
effect of the multiplicity of paths between two nodes on the infection
time. In particular, we demonstrate that the existence of even long
paths reduces the average infection time. These different results
could be of use for the design of time-dependent containment
strategies.

\end{abstract}

\maketitle

\section{Introduction}

Many complex systems display a very heterogeneous degree
distribution~\cite{Barabasi:2000,mdbook,psvbook,Newman:2003}
characterized by a power law decay of the form $P(k)\sim k^{- \gamma}$. This
form implies the absence of a characteristic scale hence the name of
``scale-free network'' (SFN)~\cite{Barabasi:1999,Amaral:2000}. Among
these networks, a certain number are of a great interest to
epidemiology~\cite{Liljeros:2001,Schneeberger:2004,Newman:2003} and it
is thus very important to understand the effect of their topology on
the spreading dynamics of a disease. One of the most relevant results
is that disease spreading does not show an endemic threshold in SFN
when the population size is infinite and $\gamma\leq
3$~\cite{Cohen:2000,Pastor:2001a,Pastor:2001b,Lloyd:2001,May:2001}. This result
means that a disease propagates very easily on a large SFN whatever
the value of its transmission probability. In addition, recent studies
showed that the presence of hubs in SFN not only facilitates the spread
of a disease but also accelerates dramatically its
outbreak~\cite{Barthelemy:2004a,Barthelemy:2005,Loecher:2005}.

The long-tailed degree distribution of SFN is the signature of the
presence of a non-negligible number of highly connected nodes. These
hubs were already identified in the epidemiological literature as
superspreaders~\cite{Hethcote:1984,May:1992}. Consequently, from a
public health point of view, studying the spreading of epidemics on
SFN is all the more appropriate.  Superspreading events affect the
basic reproductive number $R_0$---a widely used epidemiological
parameter~\cite{May:1992,Diekmann:2000}---making its estimate from
real-world data
difficult~\cite{Keeling:2000,Lipsitch:2003,Riley:2003}. As a matter of
fact, it seems that superspreading events appeared in the onset of the
recent SARS
outbreak~\cite{Lipsitch:2003,Riley:2003,Galvani:2005,Lloyd:2005} and
could be crucial for the new emergent diseases and bioterrorist
threats. Their potential threat justifies detailed studies of the
incidence of the degree distribution at the initial stage of
epidemics.

The variability plays an important role in the accuracy and the
forecasting capabilities of numerical models and has thus to be
quantified in order to assess the meaningfulness of simulations with
respect to real outbreaks~\cite{Colizza:2005}. Using a numerical
approach, we analyze the evolution of epidemics generated by different
sets of initial parameters, both for SFN and homogeneous random
networks (RN). We use the Barab\'asi-Albert model
(BA)~\cite{Barabasi:2000} for generating a SFN and the Erd\"os-Renyi
network (ER)~\cite{Erdos:1960} as a prototype for RN. Concerning the
epidemic modeling, a simple and classical approach is to consider that
individuals are only in two distinct states, infected (I) or
susceptible (S). There is initially a number of $i_0~N$ infected
individuals and any infected node can pass the disease to his
neighbors~\cite{May:1992,Diekmann:2000}. The probability per unit time to
transmit the disease---the spreading rate---is denoted by $\lambda$
and once a susceptible node is infected it remains in this state.
In more elaborated models, an infected individual can change its state to 
another category, for example, coming back to susceptible (SIS), or going to 
immunized or dead (SIR)~\cite{May:1992,Diekmann:2000}.
This S $\rightarrow$ I approach (SI), in
spite of its simplicity, is a good approximation at short times to
more refined models such as the SIS or SIR models. The SI model on
both SFN and RN is thus well adapted to the characterization of the
variability of the initial stages of epidemic outbreaks spreading in
complex networks, which is the focus of this article.

The outline of the paper is the following. In section II, we study the 
fluctuations of the prevalence and we identify
different parameters controlling them. In particular, we highlight the
effects due to different realizations of the network as well as
different initial conditions. We also investigate the influence of the
nodes degree on the prevalence variability. In section III, we present
results on the infection time and its variation with the degree and
with the distance from the origin of infections. We also discuss the
effect of the number of paths between two nodes on the infection
time. Finally, we discuss our results and conclude in section IV.\looseness=-1

\section{Prevalence fluctuations}
\label{sec:prev_fluct}

\subsection{Intra and inter-networks fluctuations}
\label{ssec:intra-inter-fluct}

We analyze in this section the effect of the underlying network
topology on the variability of outbreaks. It is indeed important to
understand whether the local fluctuations of the structure of the network
can have a large impact on the development of epidemics.

In order to analyze this effect, we measure the variability of
outbreaks as the relative variation of the prevalence (density of
infected individuals $i(t)$) given by
\begin{equation}
\cv[i(t)]= \frac{\sqrt{\langle i(t)^2\rangle -\langle i(t) \rangle^2}}{\langle 
i(t) \rangle}.
\end{equation}

\begin{figure}
  \psfrag{time}{\large$t$}
  \psfrag{cvit}[b]{$CV[i(t)]$}
  \includegraphics[width=0.85\columnwidth]{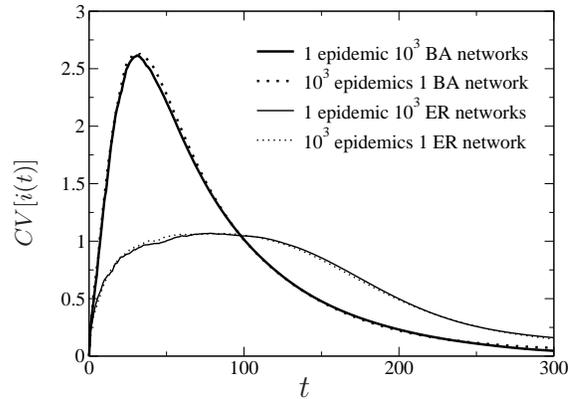}
  \caption{Evolution in time of the coefficient of variation of the
    density of infected ($\cv[i(t)]$) in BA networks (bold) and ER
    network (thin) for outbreaks simulated on the same network (dashed
    curve), or on different networks (plain curve).The results are
    obtained for $\lambda \eq 0.01$ and on networks of size $N \eq
    10^4$~nodes, and average degree $\langle k\rangle = 6$. }
  \label{fig:CV_1vs1000}
\end{figure}

In order to evaluate this quantity we run simulations for different
``model sets'': first, for a given number of outbreaks on a single
network, second for a single outbreak on different networks, and
finally several outbreaks on different networks. We show in
Fig.~\ref{fig:CV_1vs1000} the curves $\cv[i(t)]$ computed for both the
RN (thin lines) and the SFN (bold lines) and for two of these model
sets: $10^3$ outbreaks spreading on the same network (dashed curves),
and a single outbreak per network for $10^3$ different networks (plain
curves). The curves representing these two model sets are nearly
superimposed for both network topologies. The curves obtained from
model sets made of $10$ outbreaks on $100$ networks and $100$
outbreaks on $10$ networks coincide with the other model sets (not
shown in the figure). These results indicate that the contribution to
the variability of $i(t)$ given by a particular network realization is
essentially the same as the one generated by different outbreaks on
the same network. This confirms the intuitive idea that sampling
different parts of a large network is equivalent to average over
different networks. Consequently, studying variability of epidemics
simulated on one large enough network (intra-network) will lead
practically to the same conclusions as studying variability on several
instances of that network (inter-network). Furthermore, it means that
the results described in the next sections for one network can be
generalized to any instances of BA and ER networks.

\begin{figure}
\psfrag{tau}{$\large\tau=\langle k \rangle / [\lambda*(\langle k^2\rangle 
-\langle k \rangle)]$}
\psfrag{tauv}{\large$\tau_v$}
  \includegraphics[width=0.9\columnwidth]{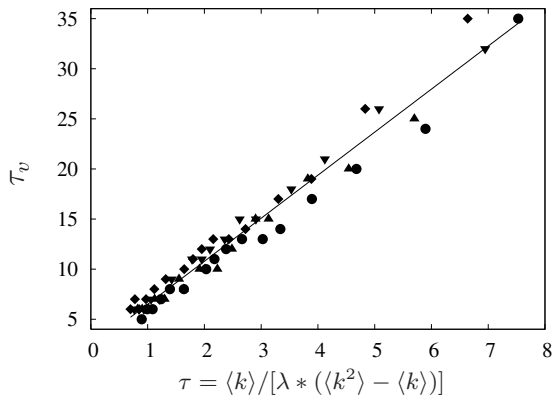}
  \caption{$\tau_v$ versus $\tau$ for several BA networks with
    $\langle k \rangle$ ranging from $6$ to $60$, and different sizes
    ($\bullet$: $N \eq 5.10^3$, $\blacktriangle$: $N \eq 10^4$,
    $\blacktriangledown$: $N \eq 2.10^4$, $\blacklozenge$: $N \eq
    5.10^4$ nodes; $\lambda=0.01$). The line is a linear fit with
    slope of order $4$.}
\label{fig:tau_v_vs_tau}
\end{figure}

Fig.~\ref{fig:CV_1vs1000} also reveals interesting facts about the
time behavior of $\cv [i(t)]$ on complex networks. Since the initial
prevalence is fixed and is the same for all instances, $\cv$ is
initially equal to zero and can only increase. At very large times,
almost all nodes are infected implying that
$\lim_{t\to\infty}\cv=0$. This argument implies the existence of a
peak which---as shown in Fig.~\ref{fig:CV_1vs1000}---is located for
BA networks at the beginning of the outbreak, with a maximum value
larger than the one obtained for ER networks.  In order to
characterize the relation between the variability peak and the network
heterogeneity, we define $\tau_v$ as the time at which the maximum of
$\cv[i(t)]$ is reached. We also use the fact that the heterogeneity of
the network degree---often quantified by $\kappa = \langle k^2
\rangle /\langle k \rangle $---is related to the typical outbreak
timescale $\tau$ given by~\cite{Barthelemy:2004a,Barthelemy:2005}
\begin{equation}
\tau = \frac{1}{\lambda(\kappa-1)}.
\label{eq:tau}
\end{equation}
A discussion of the validity of this equation is provided in 
Ref.~\cite{Vazquez:2006}.
In order to understand to which regime corresponds $\tau_v$, we plot
in Fig.~\ref{fig:tau_v_vs_tau} $\tau_v$ and $\tau$ for BA networks
with different values of $\kappa$. We use networks with different
sizes (from $N=5.10^3$ to $N=5.10^4$ nodes) and with different values
of $ \langle k\rangle$ ($6 < \langle k \rangle < 60$) in order to
obtain a broad range of $\tau$ values.

We see in Fig.~\ref{fig:tau_v_vs_tau} that $\tau_v$ is increasing
linearly with $\tau$ (with a pre-factor of order $4$).  This implies
that $\tau_v$ is of the same order of the typical time $\tau_0$ where
the diversity of degree classes of infected nodes is the largest
($\tau_0\approx 6~\tau$)~\cite{Barthelemy:2004a,Barthelemy:2005}. The
result $\tau_v\approx\tau_0$ confirms the intuitive idea that the
variability is maximal when the diversity of different classes of
infected nodes is the largest, which happens at the beginning of the
spread.

\subsection{Effect of degree on $i(t)$ fluctuations}
\label{ssec:effect_of_parameters}

\subsubsection{Seed degree}

\begin{figure}
  \psfrag{time}{\large$t$}
  \psfrag{cvit}[]{$CV[i(t)]$}
  \psfrag{it}[]{$i(t)$}
  \includegraphics[width=0.9\columnwidth]{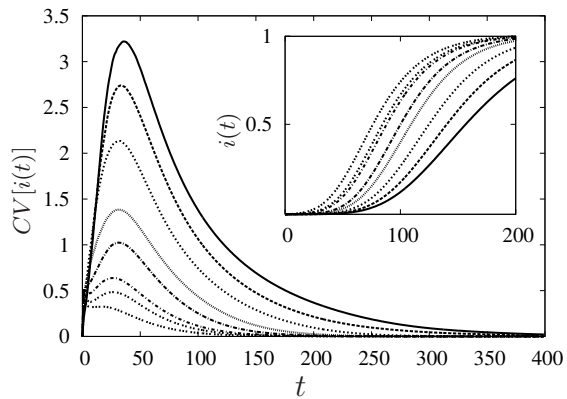}
  \caption{Temporal evolution of the coefficient of variation of the
    density of infected ($\cv[i(t)]$) in BA networks for outbreaks seeded with 
    infected nodes of different degrees $k_0$ (from top to bottom, $k_0 \eq 3, 
    6, 12, 24, 48, 95, 142, 248$).  Inset: Initial
    evolution of the prevalence i(t). The order of the curves is
    reversed between both plots (Results are averaged over
    $5.10^3$~epidemics on one network, with $\lambda \eq 0.01$, $N \eq
    10^4$ nodes, $\langle k \rangle \eq 6$).}
  \label{fig:CV_graine}
\end{figure}

In this SI model, the parameter $\lambda$ simply fixes the time
unit. In contrast, we expect that other parameters such as the degree
of the seed may have a more interesting effect on the outbreak and its
variability.  Fig.~\ref{fig:CV_graine} displays the evolution of $\cv
[i(t)]$ for outbreaks starting from initial infected nodes with a
given degree $k_0$ (from $3$ up to $248$). This figure shows that the
variability peak decreases when $k_0$ is increased.  In other words,
when an outbreak begins from a highly connected node, the early stages
of the spreading tend to be less variable. One might think that the
number of paths available on a highly connected node leads to a higher
overall variability, it is however not the case. As shown in the inset
of Fig.~\ref{fig:CV_graine}, the prevalence increases with the seed
degree, which may explain the variability for different $k_0$.
Indeed, when the seed is a hub, the number of infected becomes rapidly
very large and thus leads to smaller relative variations of the
prevalence. This result leads us to investigate more thoroughly the
degree of infected nodes and analyze the differences between BA and
ER networks.

\subsubsection{Degree of infected nodes}

In this section, we study in detail the degree properties of the
infected nodes during the  outbreak of the disease.

\begin{figure}
  \begin{tabular}{lc}
    \raisebox{5.6cm}{ { \large a)}} &
  \psfrag{t}{\large$t$}
  \psfrag{k}[B]{\large$k~$}
    \includegraphics[width=0.9\columnwidth]{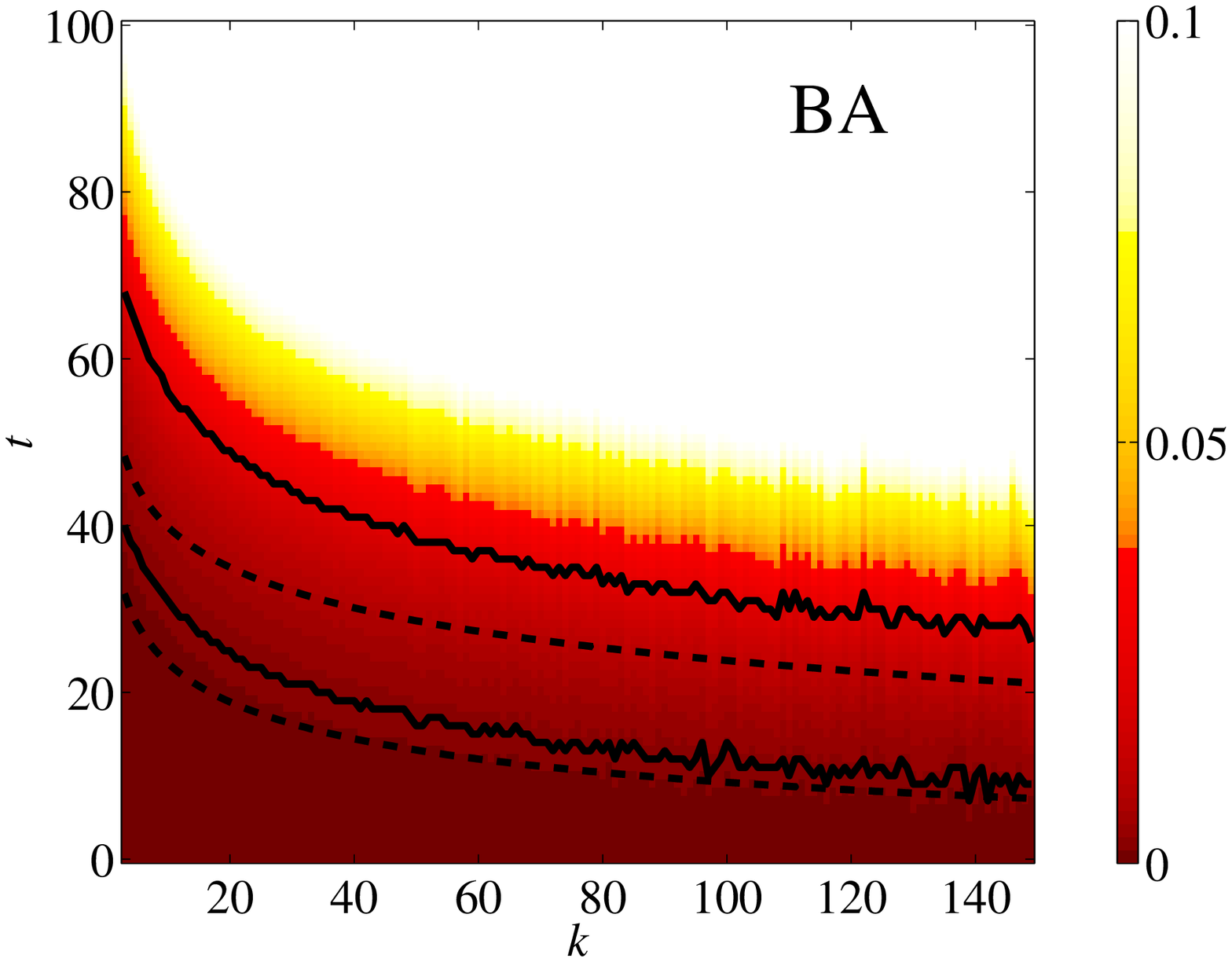}\\
    \raisebox{5.6cm}{{ \large b)}} &
  \psfrag{t}{\large$t$}
  \psfrag{k}{\large$k$}
    \includegraphics[width=0.9\columnwidth]{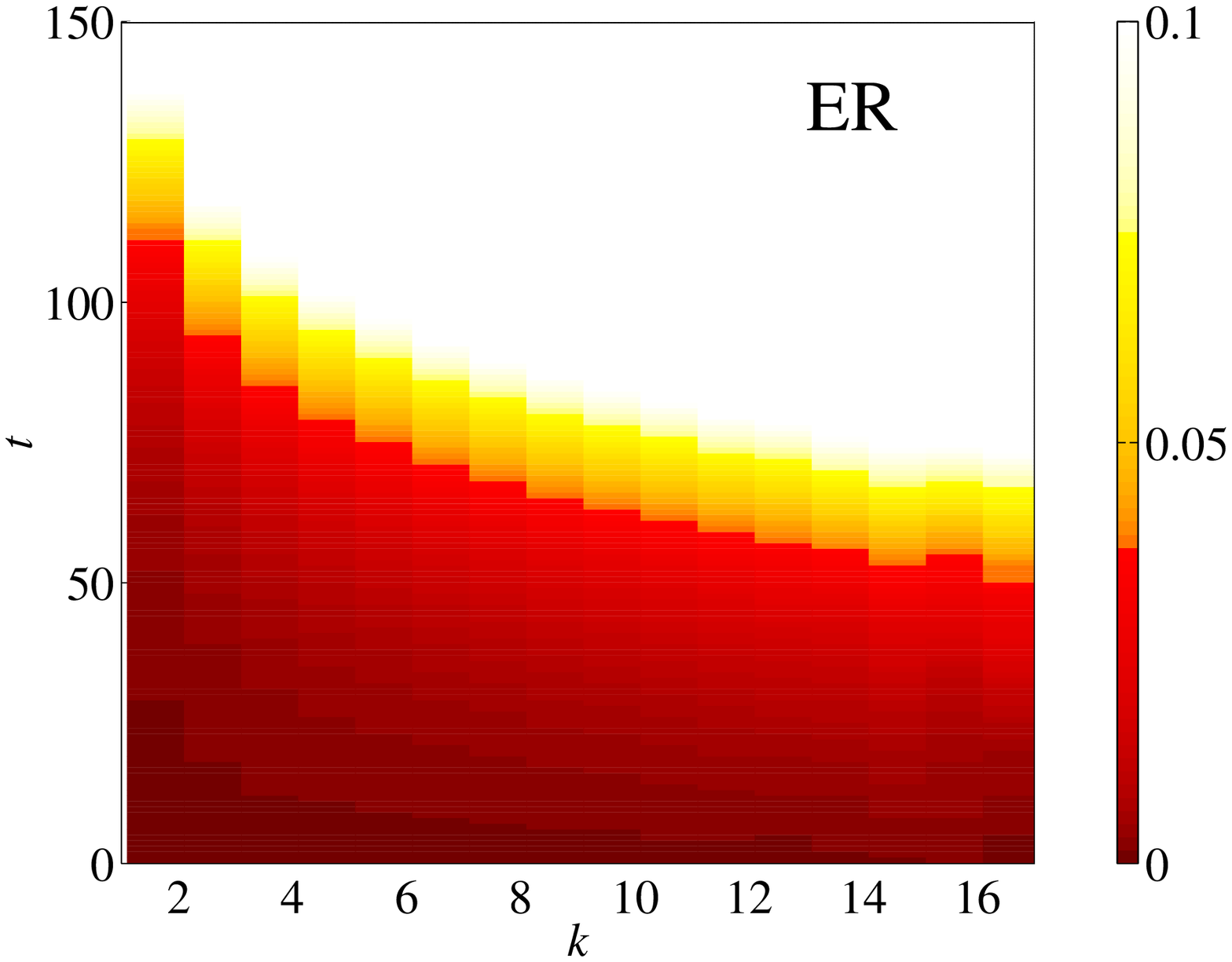}\\
  \end{tabular}
  \caption{Temporal evolution of the density of infection by classes
    of degree. {\bf (a)} Spreading on a BA network. The dashed lines
    are given by Eq.~(\ref{eq:tkg}) with different values for $i$:
    0.002, 0.02 (lower to upper) and the plain lines corresponds to
    their numerical results.  {\bf (b)} Spreading on an ER network. For
    both panels the color bar represents the density of infected and
    where white means 0.1 and above (The results are computed over
    $10^3$ outbreaks on networks of size $N \eq 10^4$ nodes, $\langle
    k\rangle \eq 6$, and spreading rate $\lambda \eq 0.01$).}
  \label{fig:dist_conn_inf}
\end{figure}

For a SI model, the evolution of the density $i_k(t)$ of infected 
nodes of degree $k$ is given at the mean-field level by
\begin{equation}        
\frac{di_k(t)}{dt}=\lambda \; k \; [1-i_k(t)] \; \Theta_k(t)
\label{eq:di_k_over_dt}
\end{equation}
where $1-i_k$ is the density of susceptible nodes of degree $k$ and
$\Theta_k$ is the probability that a link pointing to a node of degree
$k$ originates at an infected node~\cite{Pastor:2001a}.  This
equation, studied for an uncorrelated scale-free network and uniform
initial conditions $i_0 \equiv i_k(t \eq 0)$ leads to the following
behavior at short times~\cite{Barthelemy:2004a,Barthelemy:2005}
\begin{equation}        
i_k(t) \simeq i_0 \left[ 1 + \frac{k\;\langle k\rangle}{\langle k^2 \rangle 
-\langle k \rangle}(e^{t/\tau} 
- 1)\right]
\label{eq:i_k}
\end{equation}
with $\tau$ defined in Eq.~(\ref{eq:tau}). 

From this equation, we can deduce the expression for the time $t_k(i)$
for $i_k$ to reach the value $i$:
\begin{equation}        
t_k(i) 
\simeq \tau \log \left[
1 + \frac{\langle k^2 \rangle - \langle k \rangle}
{k\langle k \rangle}
\left(\frac{i}{i_0} - 1 \right)
\right]
\label{eq:tkg}
\end{equation}
For a fixed prevalence $i$, the time $t_k(i)$ varies very slowly with
$k$ and thus can vary significantly only on a network with a large
range of degree variation. The results are shown on
Fig.~\ref{fig:dist_conn_inf}, which is composed of two contour maps of
the temporal evolution of $i_k$ in both BA and ER networks. In order
to simplify the reading of this figure, the density of infection has
been limited to $0.1$ since we are only interested in the beginning of
the outbreaks. We also plot, in Fig.~\ref{fig:dist_conn_inf}(a), the
curves corresponding to Eq.~(\ref{eq:tkg}) for different values of
$i$~(0.002 and 0.02) and numerical result for the same values (plain
curves). It can be seen that the predictions of Eq.~(\ref{eq:tkg}) for
small density and short times are in agreement with the average
behavior obtained from our simulations (the agreement is better for
larger degree since the hubs are infected at smaller times).  For
larger times, the approximation used in Eq.~(\ref{eq:i_k}) is not
valid anymore, which explains the observed discrepancy for larger
values of the density such as $i = 0.02$. These results confirm earlier
work~\cite{Barthelemy:2004a,Barthelemy:2005} on the ``cascading
effect'' of the spreading, from hubs to poorly connected
nodes. Figure~\ref{fig:dist_conn_inf}(b) is the ER counterpart of
Fig.~\ref{fig:dist_conn_inf}(a). It demonstrates that the hierarchical
spreading from well connected to poorly connected nodes also occurs on
homogeneous networks. The cascading effect however is less visible on
the average degree of infected nodes because of the limited range of
degrees (see also Sec.~\ref{ssec:eff_conn_tinf_fluct}).

\begin{figure}
  \psfrag{time}{\large$t$}
  \psfrag{k}[B]{\large$k$}
    \includegraphics[width=0.9\columnwidth]{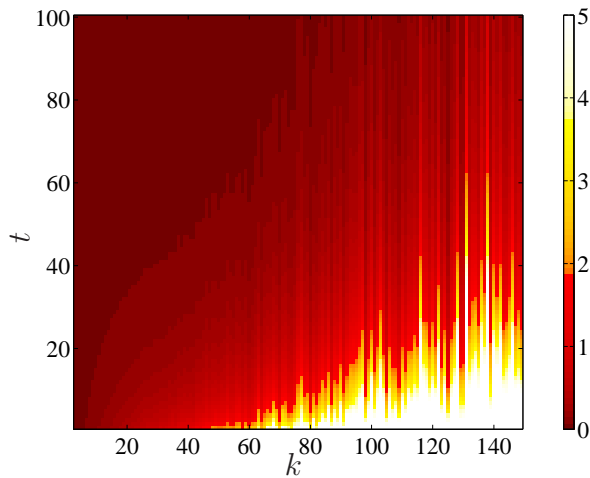}
    \caption{Temporal evolution of the coefficient of variation of the
    density of infected by classes of degree on a BA network. The
    range of $\cv[i_k(t)]$ is limited to $[0,5]$ for readability (the
    actual results can go up to $30$). Color bar accounts for
    $\cv[i_k(t)]$ (white means 10 and above). These results are
    obtained for $N \eq 10^4$ nodes, $\langle k\rangle \eq 6$,
    $\lambda \eq 0.01$, and averaged over $10^3$ outbreaks on $50$
    different networks in order to have data for the whole range of
    degrees.}
    \label{fig:cv_dist_conn_inf}
\end{figure}

Figure~\ref{fig:cv_dist_conn_inf} gives a complete picture of the
variability of $i_k(t)$ in an heterogeneous network and helps to
understand the role of each degree in the variability peak observed in
Fig.~\ref{fig:CV_1vs1000}. It displays for a BA network a contour map
representation of the temporal evolution of $\cv[i(t)]$ according to
the classes of degree. We observe that the largest values of
$\cv[i_k(t)]$ are reached at the beginning of outbreaks, then decrease
during the infection process. The very high values of $\cv$ (white on
the plot) which can be up to $30$ are reached during a period lasting
until $6~\tau$ (in this plot $\tau \approx 7$). The end of this period
corresponds to the moment when all degree classes are infected. For
superspreaders, Fig.~\ref{fig:cv_dist_conn_inf} also shows that their
infection time is fluctuating a lot even for long times, because of
their small number in networks.  This result will be confirmed in the
next section and means that their infection time has important
fluctuations. For some outbreaks, the time to reach a superspreader
can be long because of its distance to the seed (see
Sec.~\ref{ssec:eff_dist_tinf_fluct}).

\section{Fluctuations of the infection time}
\label{sec:inf_time_fluct}

The randomness of the epidemic process makes it very difficult to
predict an accurate time interval for the infection of a given
node. However, with the same methods used in the previous section, we
can draw the general picture of the distributions of the infection time
$t_{inf}$---defined as the time for which a given node becomes
infected---as a function of the degree of the node and its distance
to the seed (similar considerations were studied
in~\cite{Loecher:2005}).

\subsection{Effect of the degree}
\label{ssec:eff_conn_tinf_fluct}

\begin{figure}
  \begin{tabular}{lc}
    \raisebox{5.4cm}{\large a)} &
  \psfrag{tinf}{\large$t_{inf}$}
  \psfrag{k}[B]{\large$k$}
    \includegraphics[width=0.9\columnwidth]{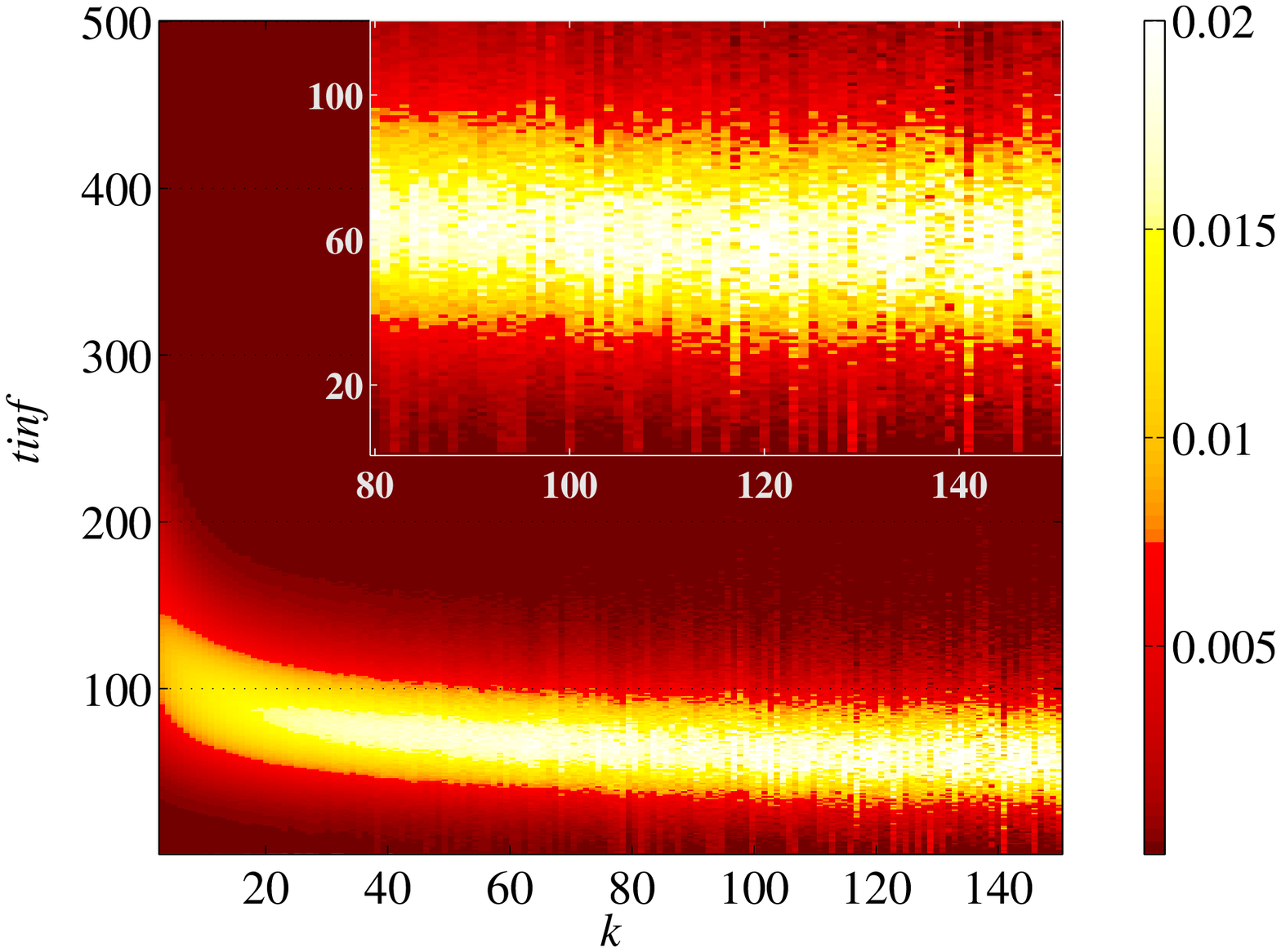}\\
    \raisebox{5.4cm}{\large b)} &
  \psfrag{tinf}{\large$t_{inf}$}
  \psfrag{k}[B]{\large$k$}
    \includegraphics[width=0.9\columnwidth]{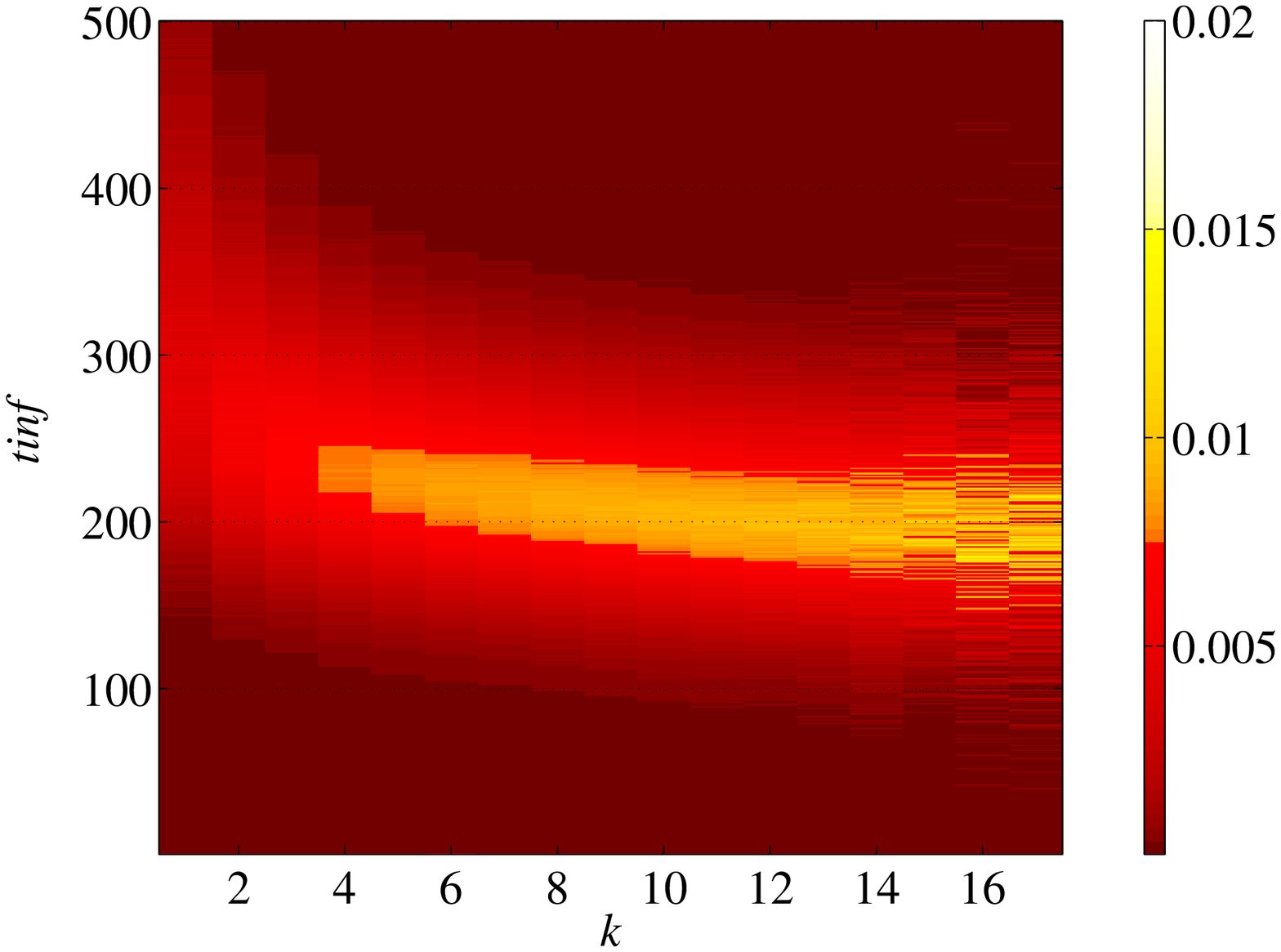}\\
  \end{tabular}
  \caption{Frequency of the moment of infection of a node as a function of
    its degree. {\bf (a)} spreading on a BA network. Inset:
    frequency of $t_{inf}$ shown for the beginning of outbreaks on
    high degree nodes. {\bf (b)} spreading on an ER network.  For both
    panels the color bar represents the frequency and where white
    means $0.02$ and above. ($N \eq 10^4$ nodes, $\langle k\rangle \eq
    6$, $\lambda \eq 0.01$, $\tau_{_{BA}} \approx 7$, $\tau_{_{ER}}
    \approx 16.5$).  }
  \label{fig:density_seq_degree}
\end{figure}

Fig.~\ref{fig:dist_conn_inf} shows how the prevalence $i(t)$ varies
with the degree. Time of infection and prevalence being related, we
first plot (Fig.~\ref{fig:density_seq_degree}) the distribution of the
infection time $t_{inf}$ versus the degree. For this figure we count
all the nodes with a given degree $k$ which have been infected at each
instant $t$, and then we normalize the corresponding results by the
number of individuals with degree $k$ and by the number of
simulations. Each degree is represented by a column where frequencies
are associated with a representative color (right color bar), the sum
of all frequencies in a column being equal to one. Given that a single
BA network does not contain the whole range of degrees, the plot shown
on Fig.~\ref{fig:density_seq_degree}(a) is based on data from 50
networks.  These results are a consequence of the cascading effect on
lower degree nodes on both topologies: the larger the degree and the
smaller the average infection time. In addition, we observe that there
is a relatively large range of fluctuation of the infection time even
for large degrees. Indeed, in the inset of
Fig.~\ref{fig:density_seq_degree}(a) we observe that for highly
connected nodes (e.g. from $80$ to $150$), the typical $t_{inf}$
varies between $6~\tau$ and $13~\tau$ (on the plot, $t= 40$ and $90$)
which is late for well-connected nodes. In fact, only a small fraction
of the superspreaders is infected during the early epidemic stages
(until $6~\tau$) and triggers the outbreak. Approximately the same
scenario seems to hold for ER networks
(Fig.~\ref{fig:density_seq_degree}(b)), even if the concept of
superspreaders is not the most appropriate for a network with a small range of
degree variation.

\begin{figure}
  \psfrag{k}[B]{\large$k$}
  \psfrag{cvtinf}[b]{$\cv(t_{inf})$}
  \includegraphics[width=0.9\columnwidth]{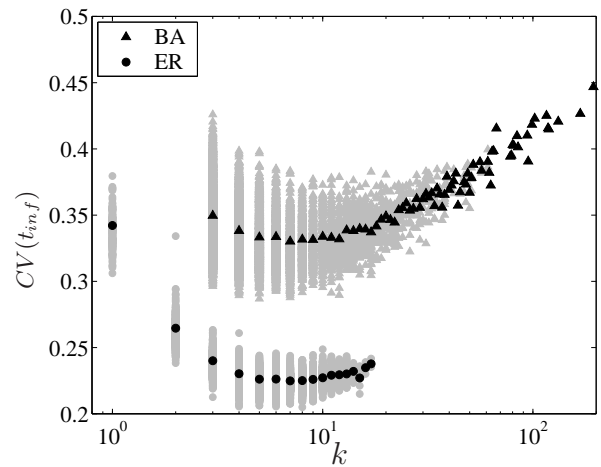}\\
  \caption{Coefficient of variation of infection time as a function of
    nodes degree $k$. Gray symbols stand for $\cv(t_{inf})$ computed
    by nodes, and black symbols for $\cv(t_{inf})$ computed for nodes
    with the same degree (vertically aligned).  $\blacktriangle$ symbols
    represent the spread on BA networks, and $\bullet$ stand for ER
    networks (Results are computed over $10^3$ outbreaks on a single
    network , $N \eq 10^4$ nodes, $\langle k\rangle \eq 6$ links, with
    a seed of degree $k_0 \eq 6$, $\lambda \eq 0.01$).  }
  \label{fig:cv(t_inf)_vs_k}
\end{figure}

In order to understand thoroughly the properties of the infection
time, we also show in Fig.~\ref{fig:cv(t_inf)_vs_k} scatter-plots of
its relative dispersion $\cv(t_{inf})$ versus the degree for both ER
and BA topologies. This figure displays more insights concerning the
behavior of $t_{inf}$ depicted in
Fig.~\ref{fig:density_seq_degree}. First, for the BA network, nodes
with a given degree $k$ can have a wide range of $\cv(t_{inf})$ which
increases with $k$. This demonstrates that even if the superspreaders
are infected at relatively short times, large relative fluctuations
cannot be excluded.  In contrast, all nodes for the ER network have smaller and 
similar values of $\cv(t_{inf})$ which is consistent with the fact
that the hierarchical spreading is less pronounced on ER due to its
limited range of degree.

\subsection{Effect of distance}
\label{ssec:eff_dist_tinf_fluct}

Another important parameter which affects the infection time
of a node is its distance to the seed as measured by the number of
hops of the shortest path~\cite{Loecher:2005}. In the networks
considered here there is no spatial component and the distance
between two nodes is given by the smallest number of hops $\ell$ to go
from one node to another.
\begin{figure}
  \psfrag{ell}{\large$\ell$}
  \psfrag{avgtinf}{\large$\langle t_{inf} \rangle$}
  \includegraphics[width=0.9\columnwidth]{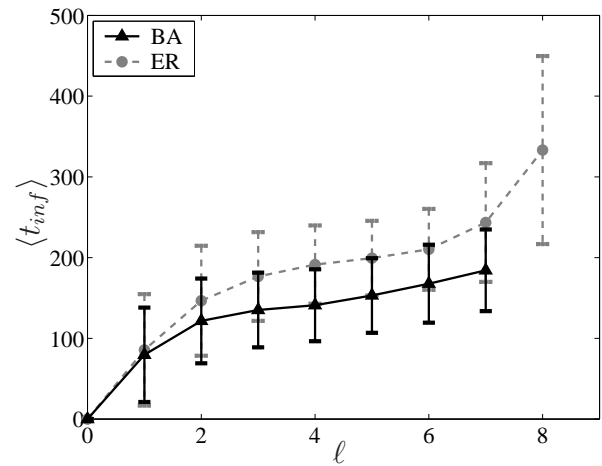}
  \caption{Infection time as a function of the distance $\ell$ from
    the seed ($N=10^4$, $\langle k \rangle = 6$, averaged over $10^3$
    outbreaks which start at exactly the same seed of degree
    $k_0=6$).  }
    \label{fig:density_seq_dtopo_sfn_rn}
\end{figure}
On Fig.~\ref{fig:density_seq_dtopo_sfn_rn}, we show the relationship
between the average time of infection $\langle t_{inf} \rangle$ and
$\ell$ for ER and BA networks. We see on this plot that the
infection time $\langle t_{inf} \rangle$ is always larger for ER
than for BA networks. It means that nodes with the same value of
$\ell$, i.e.  at the same distance from the first infected node, have
a lower $\langle t_{inf} \rangle$ if they belong to a BA networks.
\begin{table}
  \centering
\begin{tabular}{ccc}
\hline
\begin{minipage}{2cm}
${\bf\ell}$
\end{minipage}
&
\begin{minipage}{3cm}
{\bf BA networks}
\end{minipage}
& 
\begin{minipage}{3cm}
{\bf ER networks}
\end{minipage}\\
\hline
1&       1&                 1\\
2&       1.00668&      1.00144\\
3&       1.09870&      1.01126\\
4&       1.48679&      1.06732\\
5&       2.44046&      1.40124\\
6&       3.25166&      2.77646\\
7&       3.15678&      2.64469\\
8&       {--}&	       1.11256\\
9&       {--}&	       1\\
\end{tabular}
  \caption{Average number of shortest paths between a randomly selected node 
  and a node at distance $\ell$ (results are computed over $10^3$ random 
  selections of an initial node of degree $k_0=6$). $N=10^4$ nodes, $\langle k 
  \rangle = 6$.}
\label{tab:nb_sp}
\end{table}
The reason for this behavior lies in the difference of the numbers of
shortest paths in these networks. Indeed, if we enumerate these paths, we
observe that their numbers relatively differ between both
BA and ER topologies. We have computed the size and the number of shortest paths
between a randomly selected node, i.e. a potential seed of infection
and the rest of the network and we present in Table~\ref{tab:nb_sp}
the average number of shortest path at distance $\ell$. Results are
computed over $10^3$ random selection of the potential seed in order
to get an accurate picture of the network. The table exhibits a
difference in the number of path for $\ell > 2$ ($8\%$ difference for $\ell = 3$,
$40\%$ for $\ell = 4$, $74\%$ for $\ell=5$) which confirms the fact
that on BA networks, nodes have more paths to go from one to another
in a small number of hops.
\begin{figure}[h]
\begin{center}
\psfrag{avg_t}{\large$\langle t_{inf} \rangle$}
\psfrag{A}{$A$}
\psfrag{B}{$B$}
\psfrag{A}{$A$}
\psfrag{B}{$B$}
\psfrag{C}{$C$}
\psfrag{lambda}{\large$\lambda$}
\includegraphics[width=0.9\columnwidth]{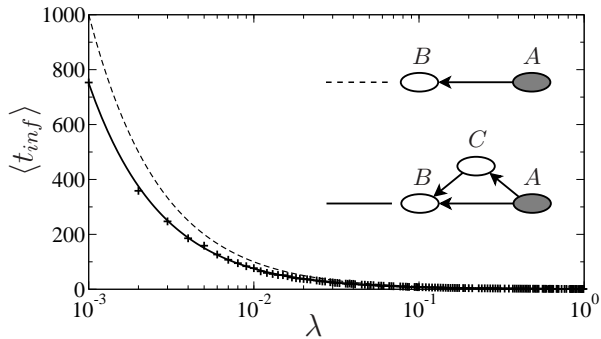}
\caption{Average infection time of node B as a function of $\lambda$
for two different configurations. In the first case infection occurs
in one step and in the second case another path is added. The dotted
curve represents the average time of infection for the first case,
$\langle t_{inf}\rangle=1/\lambda$ and the plain curve represents
$\langle t_{inf}\rangle$ for the second case and is given by
Eq.~(\ref{eq:mintdti}). The result of a numerical simulation are shown by
$+$ symbols).}
\label{fig:avg_t}
\end{center}
\end{figure}
Table~\ref{tab:nb_sp} describes the statistics of shortest paths but
longer paths also contributes to the spreading of the disease. Their
role can be highlighted by studying the following simple cases. In the
first case an infected node $A$ is in contact with a susceptible node
$B$. In the second case, there is an additional path from $A$ to $B$
going through a susceptible node $C$ (see Fig.~\ref{fig:avg_t}). In the first
``direct'' case, the average time of infection $\langle
t_{inf}^{d}\rangle(B)$ of $B$ is given by
\begin{equation}
\langle t_{inf}^{d}(B) \rangle = \frac{1}{\lambda} .
\label{eq:td}
\end{equation}
The addition of a longer path in the second case
(Fig.~\ref{fig:avg_t}) changes the behavior of $\langle
t_{inf}(B)\rangle$ and Eq.~(\ref{eq:td}) no longer holds for this
case. In fact, the time of infection of the susceptible node $B$ is
given by
\begin{equation} t_{inf}(B) = \min[t_{inf}^{d}(B), t_{inf}^{i}(B)],
\label{eq:mintdti}
\end{equation}
where $t_{inf}^{d}(B)$ is the time of a direct infection $A\to B$ and $t_{inf}^{i}$ of an 
indirect 2-steps infection process: $A\to C\to B$. The statistics of $t_{inf}$ can be easily 
computed and its first moment reads
\begin{equation}
\langle t_{inf} (B)\rangle = \frac{1}{\lambda}\frac{3 - 2 \lambda}{(2 - 
\lambda)^2}
\label{eq:avg_t}
\end{equation}
Eq.~(\ref{eq:avg_t}) predicts values always smaller than $1/\lambda$
(see Fig.~\ref{fig:avg_t}). This result could appear as paradoxical
since adding a {\it longer} path actually {\it reduces} the average
infection time. In fact, the probability that the disease is not
transmitted on both paths is very small and the existence of another
path cuts off large direct infection time and thus reduces the average
infection time of $B$. Since BA networks have a clustering coefficient
larger than ER networks~\cite{Barabasi:2000} this result explains the
small difference of infection times for $\ell=1$ seen in
Fig.~\ref{fig:density_seq_dtopo_sfn_rn}.

\begin{figure}
  \psfrag{ell}{\large$\ell$}
  \includegraphics[width=0.9\columnwidth]{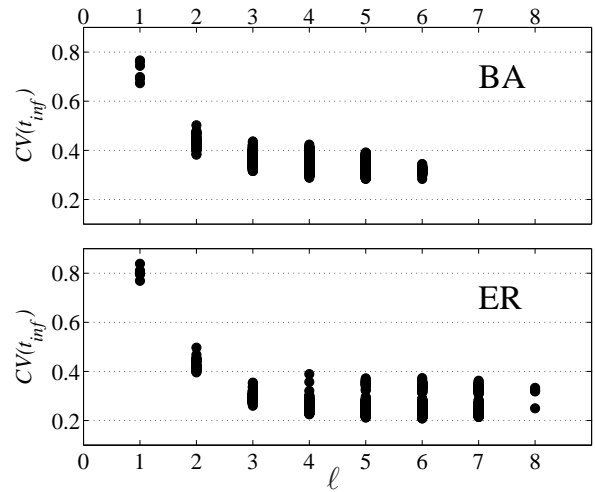}
  \caption{Coefficient of variation of the infection time as a
    function of the distance $\ell$ from the seed. Top panel:
    spreading on a BA network; bottom panel: ER network.
    Both panels show $\cv(t_{inf})$, for every nodes of a single
    network, $N=10^4$ nodes, $\langle k \rangle = 6$, and
    computed over $10^3$ outbreaks, $\lambda=0.01$, originating from
    exactly the same seed of degree $k_0=6$.  }
    \label{fig:ell_cv}
\end{figure}

Concerning the relationships between the relative dispersion of
infection time $\cv(t_{inf})$ and $\ell$, their behavior on both
topologies are reported on Fig.~\ref{fig:ell_cv}. This figure shows
that the nodes in both networks exhibit higher values of
$\cv(t_{inf})$ when they are closer to the seed, i.e.  for $\ell <
3$. For larger distances, $\cv(t_{inf})$ is practically constant in both cases.

\section{Conclusions}

We have analyzed in detail the variability of a simple epidemic
process on SFN. First, we have shown that different realizations of
BA networks do not display significant statistical differences in
outbreak variability. Consequently, it is statistically reliable to
consider a single realization of the network, provided it is large
enough. We have also shown that the prevalence fluctuations are
maximal during the time regime for which the diversity of the degrees
of the infected node is the largest. In order to analyze in detail
this variability, we examined the temporal degree pattern of infected
nodes. In particular, we demonstrated the high variability of
superspreaders' prevalence. We found that for the hubs
the infection time is usually small but with fluctuations which can be
large.  Even if the hubs are good candidates for being chosen as surveillance 
stations---given their short average infection time, they present
non-negligible fluctuations which limit their reliability. In this respect,
the ideal detection stations should be nodes with the best trade-off
between a short average infection time and a high reliability as given
by small infection time fluctuations.

The topological distance to the seed is also an important parameter in
epidemic spreading pattern. Nodes at a short distance from the seed
are infected at small time---in the high variability regime---and thus
have a large infection time variability. Maybe more surprising is the
importance of the number of paths---not only the shortest one---going
from the seed to another node. The larger this number and the smaller
the average infection time. This is an important conclusion for
containment strategies since the reduction of epidemic channels will
increase the delay of the infection arrival and will thus allow for a
better preparation against the disease (for example vaccination).

These results could be helpful in designing early detection and
containment strategies in more involved models which go beyond
topology and which include additional features such as passenger
traffic in airlines or city
populations~\cite{Flahault:1991,Hufnagel:2004,Colizza:2005,Eubank:2004}.


\begin{acknowledgments}

The authors thank A.-J.~Valleron for his support during this work.
P.C. acknowledges financial support from A.C.I. Syst{\`e}mes Complexes en
Sciences Humaines et Sociales, and F.A. from FRM (Fondation pour la
Recherche M{\'e}dicale). We also thank M.~Loecher and J.~Kadtke for
sharing with us their manuscript prior to publication.

\end{acknowledgments}



\end{document}